\documentclass[preprint,aps,prc,superscriptaddress,showpacs,floatfix]{revtex4}
\usepackage{graphicx}
\usepackage{wrapfig}
\usepackage{dcolumn}
\usepackage{bm}
\usepackage{amsfonts}
\usepackage{amsmath}
\usepackage{cancel}
\usepackage{multirow}
\usepackage{color}
\usepackage{ulem}
\newcommand {\be}{\begin{equation}}
\newcommand {\ee}{\end{equation}}
\newcommand {\bea}{\begin{eqnarray}}
\newcommand {\ea}{\end{eqnarray*}}
\newcommand {\ba}{\begin{eqnarray*}}
\newcommand {\eea}{\end{eqnarray}}

\def\x{{\bf x}}
\def\y{{\bf y}}
\def\r{{\bf r}}
\def\a{{\alpha}}

\begin{document}
\title{Theoretical description of three- and four-nucleon scattering states
 using bound-state-like wave functions}

\author{A. Kievsky}
\author{M. Viviani}
\affiliation{Istituto Nazionale di Fisica Nucleare, Largo Pontecorvo 3, 56100 Pisa, Italy}
\author{L.E. Marcucci}
\affiliation{Dipartimento di Fisica, Universita' di Pisa and Istituto Nazionale di Fisica Nucleare,
 Largo Pontecorvo 3, 56100 Pisa, Italy}

\begin{abstract}
Bound-state-like wave functions are used to determine the scattering matrix
corresponding to low energy $N-d$ and $p-^3$He collisions. To this end,
the coupled channel form of the integral relations derived from the Kohn
variational principle is used. The construction of degenerate bound-state-like
wave functions belonging to the continuum spectrum of the Hamiltonian is discussed.
Examples are shown using realistic nucleon-nucleon forces. 
\end{abstract}

\maketitle

\section{Introduction}

Well established methods to treat both, bound and scattering states
in $A=3,4$ systems, are the solution of the Faddeev ($A=3$) or Faddeev-Yakubovsky
($A=4$) equations in configuration or momentum space and the hyperspherical
harmonic (HH) expansion in conjunction with the Kohn variational principle (KVP). 
These methods have proven to be of great accuracy and they have been tested through
different benchmarks~\cite{benchmark1,benchmark2}.
On the other hand, other methods
are presently used to describe bound states: for example, the Green's
Function Monte Carlo (GFMC) and No Core Shell Model (NCSM) methods
have been used in nuclei up to $A=10$ and $A=12$ respectively~\cite{gfmc,ncsm}.
Attempts to use these methods in the description of scattering
states recently appeared~\cite{gfmc5,ncsm5}.

The possibility of employing bound state techniques to describe scattering states
has always attracted particular attention~\cite{harris}. 
Recently, continuum-discretized states obtained from the stochastic variational method
have been used to study single channel $\alpha+n$ scattering scattering~\cite{suzuki1}. 
The extension to treat coupled channel scattering is given in Ref.~\cite{suzuki2}.
In those approaches, the tangent of the phase-shift results in a quotient of two numbers.
In the former the numerator and denominator are obtained from two
integral relations after projecting the Schr\"odinger equation,
whereas in the latter
the numerator results from an integral relation derived by means of the
Green's function formalism and the denominator from the
normalization of the continuum-discretized state.

Recently two integral relations have been derived from the KVP~\cite{intrel}.
It has been shown that starting from the KVP, the tangent of the phase-shift
can be expressed as a quotient where both, the numerator and the denominator,
are given as two integral relations. This is similar to what was proposed
in Ref.~\cite{harris}, however the variational character of the quotient and
its strict relation with the KVP were not recognized.
In fact, it is this characteristic that
makes possible many different and interesting applications of the integral relations.
For example, in Ref.~\cite{kiev10a},
the integral relations have been used to compute phase-shifts from
bound state like functions in the $A=2,3$ systems using 
semirealistic interactions. Both $n-d$ and $p-d$ scattering were considered. 
The latter process is of particular interest since $p-d$ scattering at low energies
has been a subject of intense investigations. Initially,
the Faddeev method has been applied mainly to the neutral $n-d$ reaction.
Applications to $p-d$ zero-energy scattering were studied in configuration space
by the Los Alamos-Iowa group using $s$-wave potentials~\cite{friar83} and
realistic forces~\cite{friar84}. In those calculations the KVP was used to
correct the first order estimate of the scattering length after solving
the Faddeev equations in which the partial wave expansion of the Coulomb
potential was truncated. Low-energy $p-d$ elastic scattering has been
studied using the pair correlated hyperspherical harmonic (PHH) 
expansion~\cite{phh,kiev01a} as well.
A benchmark between these two techniques was given in Ref.~\cite{kiev01b}.
A different way to treat the Coulomb potential in few-nucleon scattering was
proposed in Ref.~\cite{deltuva1}, based on the works of Ref.~\cite{alt},
in which the Alt-Grassberger-Sandhas equations were solved using
a screened Coulomb potential and then the scattering amplitude was obtained
after a renormalization procedure. A benchmark for elastic $p-d$ scattering
up to $65$ MeV between this technique and the PHH expansion using the KVP has been
performed~\cite{benchmark3}.

Summarizing, the description of scattering states
using very accurate methods are at present circumscribed to $A\le4$ systems.
On the other hand, accurate methods to describe bound states beyond the $A=4$ mass
system exist. Therefore the discussion of new methods to extend these approaches
to treat scattering states is of interest. In this discussion the treatment
of the Coulomb interaction cannot be neglected. In the present work we would like
to show a detailed application of the integral relations derived from the KVP
in which $A=3,4$ bound-state-like wave functions are used to compute the scattering
matrix using realistic nucleon-nucleon $NN$ potentials. In particular, we face the problem of
constructing degenerate bound state wave functions at a given energy $E$ 
belonging to the continuum spectrum of the Hamiltonian. In fact, in the $A=3$
system, the elastic scattering matrix is a $2\times 2$ matrix for $J^\pi=1/2^\pm$ and
a $3\times 3$ matrix for all the other states. This means that,
at energies below the deuteron breakup threshold, there are two (for $J^\pi=1/2^\pm$)
or three (for $J > 1/2$) scattering states, at the same energy, 
differing in their asymptotic structure. For example, in the $J^\pi=1/2^+$ state,
two different asymptotic structures exist corresponding to 
$(L,S)=(0,1/2)$ or $(2,3/2)$, being $L$  the
relative angular momentum between the deuteron and the incoming nucleon and $S$ the
total spin $S$. Therefore, at a given energy, the two scattering states have a
particular combination of the two different asymptotic structures determined
by the scattering matrix. In the present paper we discuss how to construct
degenerate bound-state wave functions at a particular energy, belonging to the
continuum spectrum of the Hamiltonian.
Moreover, these states will be used in the integral relations to compute
the scattering matrix. Examples using realistic forces in the $A=3,4$ systems will be
shown. We expect that this study will serve as a guide for calculating
scattering states in systems with $A>4$.

The paper is organized as follow. In the next section general $A=3,4$
bound-state wave functions are constructed using the HH expansion.
In Section III, a brief derivation of the KVP given in terms of the
integral relations is discussed. Applications to the $A=3,4$ systems
are shown in Section IV whereas the conclusions are given in the last section.

\section{$A=3,4$ bound like states with arbitrary $J^\pi$ values}

Following Refs.~\cite{phh,kiev93,kiev97,report} we give a brief description
of a general three- and four- nucleon bound state in terms of the hyperspherical
harmonic basis. In the case of $A=3$, a bound-state wave function can be written 
as a sum of three amplitudes
\begin{equation}
 \Psi = \psi(\x_i,\y_i)+\psi(\x_j,\y_j)+\psi(\x_k,\y_k)\ ,
\label{eq:Psi}
\end{equation}
where $\x_i,\y_i$ are the internal Jacobi coordinates which
are defined in terms of the particle coordinates as
\begin{equation}
 \x_i=\r_j-\r_k\, ,\quad \y_i={1\over {\sqrt3}}(\r_j+\r_k-2\r_i)\ .
\end{equation}
Each $i$--amplitude has total angular momentum and parity $J^\pi$ and
third component of the total isospin $T_z$.
Using $LS$ coupling, it can be decomposed into channels
\begin{eqnarray}
     \psi(\x_i,\y_i) &=& \sum_\alpha^{N_c} \phi_\alpha(x_i,y_i)
     {\cal Y}_\alpha (jk,i)  \\
     {\cal Y}_\alpha (jk,i) &=&
     \Bigl\{\bigl[ Y_{\ell_\alpha}(\hat \x_i)  Y_{L_\alpha}(\hat \y_i)
     \bigr]_{\Lambda_\alpha} \bigl [ s_\alpha^{jk} s_\alpha^i \bigr ]
     _{S_\alpha}
      \Bigr \}_{J J_z} \; \bigl [ t_\alpha^{jk} t_\alpha^i \bigr ]_{T_\alpha T_z},
\end{eqnarray}
where $x_i,y_i$ are the moduli of the Jacobi coordinates.
Each $\alpha$--channel is labeled by the angular momenta $\ell_\alpha,
L_\alpha$, coupled to $\Lambda_\alpha$, and by the spin (isospin)
$s_\alpha^{jk}$ ($t_\alpha^{jk}$) of the pair $j,k$, coupled to the spin
(isospin) of the third particle $s_\alpha^i$ ($t_\alpha^i$) to give
$S_\alpha$ ($T_\alpha$). $N_c$ is the number of channels taken into account in the
construction of the wave function and should be increased until convergence
is reached. The antisymmetrization of the state requires that
$\ell_\alpha+s_\alpha^{jk}+t_\alpha^{jk}$ be odd, while the parity of
the state is given by $\ell_\alpha+L_\alpha$.

Defining the hyperradius and hyperangle in terms of the moduli of the Jacobi
coordinates
\begin{equation}
x_i=\rho\cos\phi_i \, ,\,\,\,\,\,\,\, y_i=\rho\sin\phi_i
\end{equation}
the two--dimensional spatial amplitudes can be expanded in
terms of the PHH basis as
\begin{equation}
     \phi_\alpha(x_i,y_i) = \rho^{\ell_\alpha+L_\alpha} f_\alpha (x_i)
     \left[ \sum_K u^\alpha_K(\rho) {}^{(2)}P^{\ell_\alpha,L_\alpha}_K(\phi_i)
     \right] \ ,
\label{eq:PHH}
\end{equation}
where the hyperspherical polynomials are
\begin{equation}
    {}^{(2)}P^{\ell_\alpha,L_\alpha}_K(\phi_i)=N_n^{\ell_\alpha,L_\alpha}
    (\sin\phi_i)^{L_\alpha}(\cos\phi_i)^{\ell_\alpha}
    P_n^{L_\alpha+1/2,\ell_\alpha+1/2}(\cos2\phi_i) \ .
\end{equation}
$N_n^{\ell_\alpha,L_\alpha}$ is a normalization factor, $P_n^{\alpha,\beta}$
is a Jacobi polynomial and $K=\ell_\alpha+L_\alpha+2n$ is the grand orbital
quantum number which runs from its minimum value $K_0=\ell_\alpha+L_\alpha$
to its maximum selected value $K_\alpha$. Therefore, the number of
hyperradial functions per channel is $M_\alpha=(K_\alpha-K_0)/2+1$.
The inclusion of the pair correlation function $f_\alpha(x_i)$ in
the expansion of Eq.(\ref{eq:PHH}) accelerates the convergence taking into account
the correlations introduced by the strong repulsion of the $NN$ potential
(see for example Ref.\cite{kiev93}).

In the case of the four-nucleon system we use the HH expansion as described
in Ref.~\cite{viv05}. The wave function having total angular momentum $J$
and parity $\pi$ can be cast in the form
\begin{equation}\label{eq:PSI3}
  \Psi= \sum_{[K]}\sum_{\alpha} 
    u^\alpha_{[K]}(\rho)
    \Psi_{\alpha}^{[K]}\ ,
\end{equation}
where $[K]\equiv K,\Lambda,S,T$ and $\Psi_{\alpha}^{[K]}$ are the channel 
HH-spin-isospin functions having
grand angular momentum $K$, orbital angular momentum $\Lambda$, coupled to
total spin $S$, to give a total angular momentum $JJ_z$, and total isospin $T$.
The channel index $\alpha$ labels the possible choices of hyperangular, spin and
isospin quantum numbers, namely
\begin{equation}
   \alpha \equiv \{ \ell_1,\ell_2,\ell_3, L_2 ,n_2, n_3, S_a,S_b, T_a,T_b
   \}\ ,\label{eq:mu}
\end{equation}
compatibles with the given values of $K$, $\Lambda$, $S$,
$T$, $J$ and $\pi$. The channel function $\Psi_{\alpha}^{[K]}$
is constructed as a linear combination of the following basis
elements
\begin{equation}
  \Psi^{[K]}_{\alpha} =  \biggl \{
   {\cal Y}^{K,\Lambda,M}_{\ell_1,\ell_2,\ell_3, L_2 ,n_2, n_3}(\Omega)
      \biggl [\Bigl[\bigl[ s_1 s_2 \bigr]_{S_a}
      s_3\Bigr]_{S_b} s_4  \biggr]_{SS_z} \biggr \}_{JJ_z}
      \biggl [\Bigl[\bigl[ t_1 t_2 \bigr]_{T_a}
      t_3\Bigr]_{T_b} t_4  \biggr]_{TT_z}\ .
     \label{eq:PHI}
\end{equation}
Here, ${\cal Y}^{K,\Lambda,M}_{\ell_1,\ell_2,\ell_3, L_2 ,n_2, n_3}(\Omega)$ is the
four-nucleon HH state and $s_i$ ($t_i$) denotes the spin
(isospin) function of particle $i$ and $\Omega$ indicates the set of the four-nucleon 
hyperangular variables. The total parity of the state is given by
$\pi=(-1)^{\ell_1+\ell_2+\ell_3} $.

\begin{figure}[h]
\includegraphics[width=16cm]{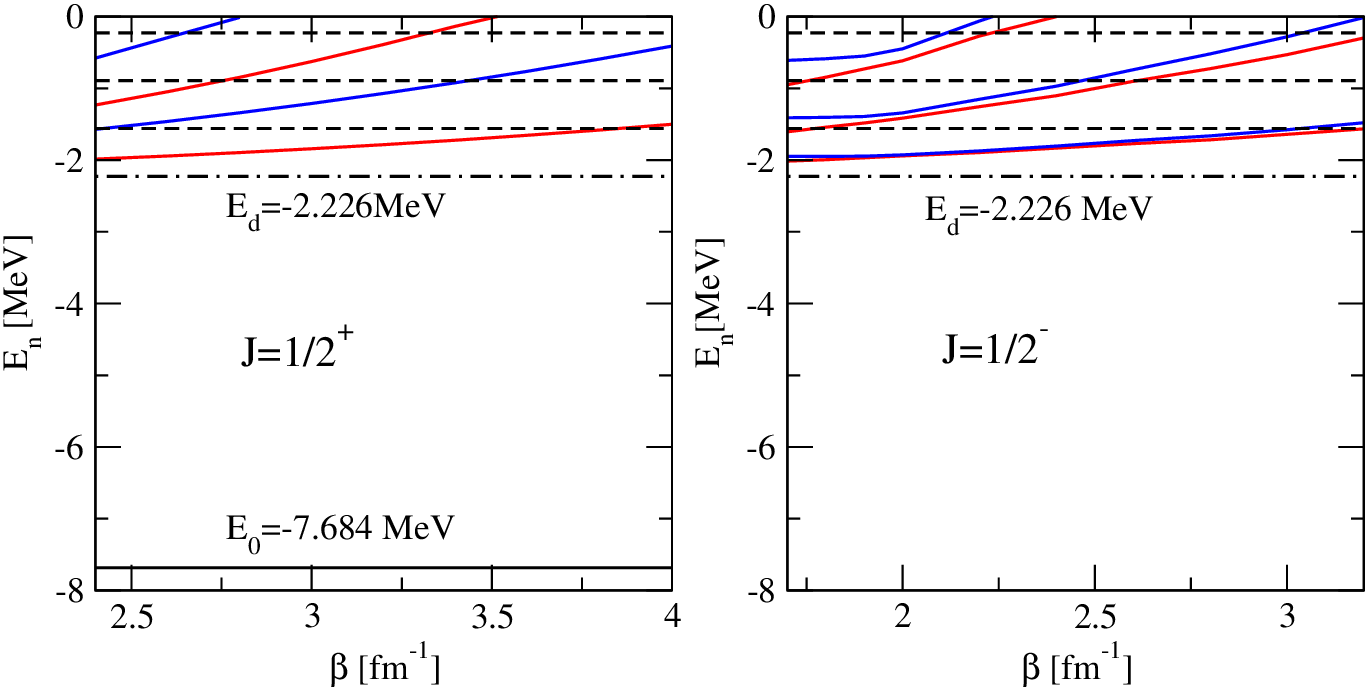}
\caption{(Color online)
The lowest $A=3$ eigenvalues, using the AV14 potential, for the $J^\pi=1/2^+$ (left panel) and
$J^\pi=1/2^-$ (right panel) states, as a function of the nonlinear parameter $\beta$.
The solid black line (left panel) represents the triton energy whereas the colored lines 
indicate the eigenvalues embedded in the continuum as explained in the text.
The dotted-dashed line represents the deuteron energy, the dashed lines indicate
the labs energies $E_{lab}=1,2,3$ MeV.}
\label{fig:fig1}
\end{figure}

In the present work the $A=3,4$ hyperradial functions $u^\alpha_{[K]}(\rho)$ 
are taken as linear combinations of Laguerre polynomials multiplied by
an exponential function:
\begin{equation}
 u^\alpha_{[K]}(\rho)=\sum_m A_{\alpha,[K],m}L^{(\gamma)}_m(z)\exp(-z/2)\ ,
\label{eq:hfun}
\end{equation}
where $A_{\alpha,[K],m}$ are coefficients to be determined and
the indeces $\alpha,[K]$ label either a three-nucleon or a four-nucleon
channel. The polynomials depend on the variable
$z=\beta\rho$, with $\beta$ a nonlinear variational parameter.
Let us define $|\alpha,[K],m>$ as a totally antisymmetric element of the expansion
basis for the $A=3,4$ systems. In terms of the basis
elements, the bound-state wave functions given in Eqs.(\ref{eq:Psi}) and 
(\ref{eq:PSI3}) can be written as
\begin{equation}
 \Psi_n=\sum_{\alpha,[K],m}A^n_{\alpha,[K],m}|\alpha,[K],m>.
\end{equation}
The index $n$ indicates the level of the state with energy $E_n$.
The linear coefficients $A^n_{\alpha,[K],m}$ of 
the wave function and the
energy of the state are obtained by solving the following
generalized eigenvalue problem
\begin{equation}
\sum_{\alpha',[K'],m'}A^n_{\alpha',[K'],m'}
    <\alpha,[K],m|H-E_n|\alpha',[K'],m'>=0 \ .
\label{eq:matrix}
\end{equation}
In the latter equation the dimension of the involved matrices
is related to three
indices: the number of $\alpha$--channels $N_c$, the number of hyperspherical
polynomials for each channel $M_\alpha$ and, $N_L$ the number of Laguerre polynomials
included in the expansion of the hyperradial functions of Eq.(\ref{eq:hfun}).
The convergence properties of the expansion is analyzed by increasing the indices $K,m$
and studying the stability obtained for different values of the nonlinear
parameter $\beta$.
The ground state of the three-nucleon system has total angular
momentum and parity $J^\pi=1/2^+$ and 
with $N_c=18$ an accuracy of $1$ keV is reached~\cite{kiev97,kame89}. The 
corresponding dimension of the PHH basis is $D\approx 2200$, considering $M_\alpha=8$ for the
first $8$ channels, $M_\alpha=6$ for the successive six channels,
$M_\alpha=4$ in the last ones and including $N_L\approx 20$ Laguerre polynomials in the
description of the hyperradial functions. After the diagonalization of the whole
matrix, $D$ eigenvalues are obtained. The lowest one corresponds to 
the three-nucleon ground state and, with the very extended basis used, it shows a 
noticeable stability with $\beta$. A
certain number of negative eigenvalues verifying $E_n>E_d$ (with $E_d$ the deuteron
energy) also appear. Defining the positive energy $E^0_n=E_n-E_d$, the corresponding
eigenvectors $\Psi_n$ approximately describe a scattering process at the center-of-mass 
energy $E^0_n$, though asymptotically they go to zero. 
The eigenvalues $E_n$ present a monotonic behavior with $\beta$, as shown
in the left panel of Fig.~\ref{fig:fig1}, where the AV14 $NN$ 
potential~\cite{av14} has been used. 

In the right panel of Fig.~\ref{fig:fig1} the lowest eigenvalues obtained from 
a diagonalization of the $J^\pi={1/2^-}$ state are shown. 
As expected this state is not bound, though several negative states appear with 
energies in the interval $E_d<E_n<0$, characterized with a monotonic
behavior with $\beta$. As before, these states approximately describe a 
scattering process at the center-of-mass energy $E^0_n$. In Fig.~\ref{fig:fig1}
the deuteron energy
is indicated by the dotted-dashed line whereas the three dashed lines indicate the
lab energies $E_{lab}=1,2,3$ MeV, respectively.
Interestingly, the energies of the $J^\pi=1/2^-$ state appear
in pairs. This can be understood noticing that the $J^\pi=1/2^-$ scattering states
are twofold degenerate at a given energy, as the scattering matrix 
has dimension of two. This degeneration arises from the two possible asymptotic configurations
in which the relative angular momentum of the deuteron and the third nucleon is $L=1$
and the total spin can take the values $S=1/2$ and $3/2$. Also the
$J^\pi=1/2^+$ state is twofold degenerate, having two possible asymptotic configuration
with the values $L=0,S=1/2$ and $L=2,S=3/2$. However, in this case, the different $L$ values
produce different contributions to the kinetic energy with the consequence that
the two degenerate states appears with a larger separation compared to the
$J^\pi=1/2^-$ case. However, this difference reduces as the basis is enlarged.

To analyze further the hypotesis that the states organize in pairs
corresponding to the two different asymptotic configurations in both
$J^\pi=1/2^{\pm}$ states, in Table~\ref{tb:tab1} the different occupation probabilities
are given. For the $J^\pi=1/2^+$ state the occupation probabilities of the
$S$- $P$ and $D$-waves, $P_S,P_P$ and $P_D$, 
have been computed. The $E_0$ level corresponds to the ground
state and the successive levels organize in mostly $S$-wave ($E_1$ and $E_3$) and
mostly $D$-wave ($E_2$ and $E_4$) states, alternatively. 
In the case of the $J^\pi=1/2^-$ state,
the occupation probabilities of the $P$-wave with total spin values $S=1/2$ and $3/2$,
$P^{1/2}_P$ and $P^{3/2}_P$, as well as $P_D$ have been computed. From the table
we can observe that the levels organize in pairs, being one of the states
mostly a $P$-wave state with $S=1/2$ and the other mostly a $P$-wave state
with $S=3/2$. This organization is indicated in Fig.~\ref{fig:fig1} with colors.
For the $J^\pi=1/2^+$ the $E_0$ level, shown as a black solid line,
is practically constant with $\beta$. The levels with high $L=0$ ($L=2$)
occupation probability are given in red (blue) respectively. 
In the case of the $J^\pi=1/2^-$,
the levels with high $P_P^{3/2}$ ($P_P^{1/2}$) probabilities are given in
red (blue) respectively. For small values of $\beta$ the spectrum tends to be
denser since, in this case, the polynomials can contain more oscillations
before the action of the exponential tail becomes significant. 
As $\beta$ increases the number
of negative eigenvalues decreases. In the case in which a bound state
exists, as in the case of the $J^\pi=1/2^+$ state, the
basis is sufficiently large to guarantee a correct description of it
as the control parameter $\beta$ is varied. 
As we will see, the wave functions $\Psi_n$ corresponding
to energy levels $E_d<E_n<0$, can be used to determine the scattering matrix at specific
energies.

\begin{figure}[h]
\includegraphics[width=16cm]{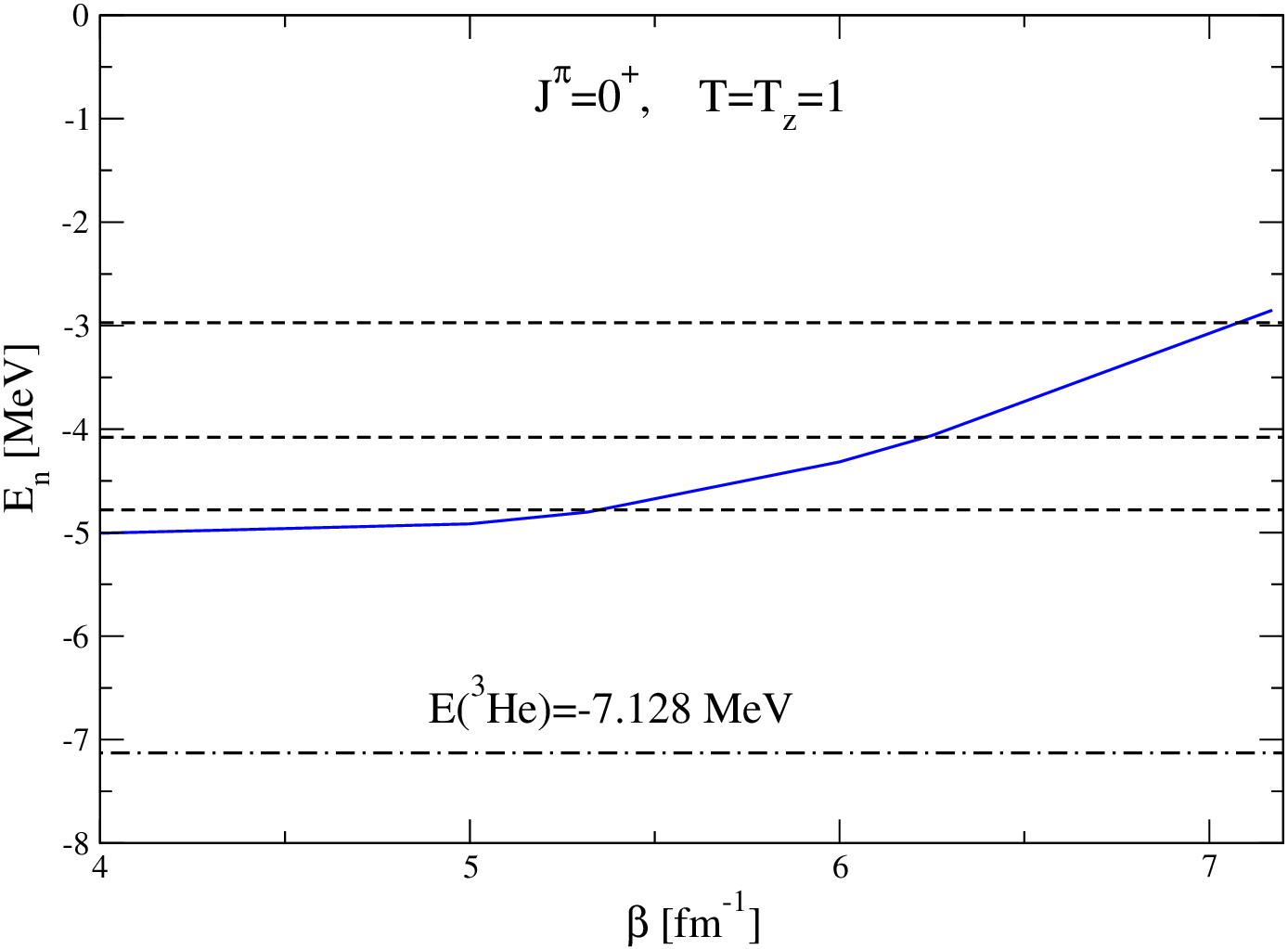}
\caption{(Color online)
The lowest $A=4$ eigenvalue, using the N3LO potential,
for the $J^\pi=0^+$ state with $T=T_z=1$ as a function
of the nonlinear parameter $\beta$. The colored lines indicate
the eigenvalues embedded in the continuum as explained in the text.
The dotted-dashed line represents the triton energy, the dashed lines indicate
the labs energies $E_{lab}=3.13,4.05,5.54$ MeV.}
\label{fig:fig4b1}
\end{figure}

In the case of the $A=4$ system we analyze the single channel
$J^\pi=0^+$ state with $T=T_z=1$, 
corresponding to the $p-^3{\rm He}$ system. 
Using the N3LO-Idaho potential~\cite{N3LO}, the Hamiltonian matrix has a total dimension
$D \approx 84000$, obtained expanding the wave function on the HH basis, as
as previously described, with $K_{max}=44$, corresponding to about $3500$ HH
states, and $N_L=24$. For this values of $D$, the matrix can be 
diagonalized using standard iterative methods. In Fig.~\ref{fig:fig4b1} the
first eigenvalue is shown as a function of the control parameter $\beta$. Clearly
the lowest eigenvalue is above the $^3$He threshold, fixed for the N3LO-Idaho potential 
at -7.128 MeV, since four nucleons in the isospin channel $T=1$ does not present
a bound state. The three dashed lines correspond to three lab energies 
(3.13, 4.05 and 5.54 MeV) at which experimental data exist. Similar to the
previous cases in $A=3$, we will use these four-body bound state wave functions to
determine the $p-^3{\rm He}$ scattering matrix at the indicated energies.

\section{The KVP in terms of integral relations}

Following Refs.~\cite{kiev01a,report} a general scattering state  with $A=3,4$
can be written as a sum of two terms
\begin{equation}
   \Psi=\Psi_C+\Psi_A \ .
\label{eq:psisc}
\end{equation}
The first term, $\Psi_C$, describes the
system when the $A$ nucleons are close to each other. For large
interparticle separations and energies below the breakup threshold in more
than two pieces it goes to zero, whereas for higher energies it must
reproduce a three or four outgoing particle state. It can be written as a 
sum of amplitudes corresponding to the cyclic permutations of the Jacobi
coordinates. Each amplitude $\Psi_C(\{\x_i\})$ has total angular momentum 
and parity $J^\pi$ and third component of the total isospin $T_z$
(here $\{\x_i\}$ represents the set of Jacobi coordinates with ordering of
the particles $i$ for the $A=3$ or $A=4$ systems). 
For energies below the breakup threshold in three pieces,
it can be expanded in terms of the totally antisymmetric states
\begin{equation}
 \Psi_C=\sum_{\alpha,[K],m}A_{\alpha,[K],m}|\alpha,[K],m>.
\end{equation}

The second term, $\Psi_A$, in the scattering wave function of
Eq.(\ref{eq:psisc}) describes the relative motion of the two clusters
in the asymptotic region. For $A=3$, $\Psi_A$ describes the relative motion
between the deuteron and the incident nucleon, whereas for $A=4$ we will
limited the description to an incident nucleon on $^3$He or $^3$H.
It can be written as a sum of amplitudes whose generic form
for $A=3$ is given by
\begin{equation}
   \Omega^\lambda_{LSJ}(\x_i,\y_i) = \sum_{l_\a=0,2} w_{l_\a}(x_i)
       {\cal R}^\lambda_L (y_i)
       \left\{\left[ [Y_{l_\a}({\hat x}_i) s_\a^{jk}]_1 s^i \right]_S
        Y_L({\hat y}_i) \right\}_{JJ_z}
       [t_\a^{jk}t^i]_{TT_z}\ , \label{eq:omega}
\end{equation}
where $w_{l_\a}(x_i)$ is the $l_\a =0,2$ deuteron wave function,
$s_\a^{jk}=1,t_\a^{jk}=0$, and $L$ is the relative angular momentum
of the deuteron and the incident nucleon. The superscript $\lambda$ indicates
the regular ($\lambda\equiv R$) or the irregular ($\lambda\equiv I$)
solution of the Schr\"odinger equation in the asymptotic region. 
In the $p-d$ ($n-d$) case, the functions
${\cal R}^\lambda$ are related to
the regular or irregular  Coulomb (spherical Bessel) functions.
The functions $\Omega^\lambda$ can be combined to form a general
asymptotic state
\begin{equation}
\Omega^+_{LSJ} =  \sum_{i=1,3}\left[ \Omega^0_{LSJ}(\x_i,\y_i)+
 \sum_{L'S'}{}^J{\cal L}^{SS'}_{LL'}\Omega^1_{L'S'J}(\x_i,\y_i)\right]  \ ,
\label{eq:scata}
\end{equation}
where
\begin{eqnarray}
\Omega^0_{LSJ}(\x_i,\y_i) =& u_{00}\Omega^R_{LSJ}(\x_i,\y_i)+
                            u_{01}\Omega^I_{LSJ}(\x_i,\y_i) \ , \\
\Omega^1_{LSJ}(\x_i,\y_i) =& u_{10}\Omega^R_{LSJ}(\x_i,\y_i)+
                            u_{11}\Omega^I_{LSJ}(\x_i,\y_i)  \ .
\end{eqnarray}
The matrix elements $u_{ij}$ form a matrix $u$ that can be selected according to the
four different choices of the matrix ${\cal L}=$ ${\cal K}$-matrix,
${\cal K}^{-1}$-matrix, $S$-matrix or $T$-matrix. It should be noticed that the irregular
solution has been opportunely regularized at the origin
\begin{equation}
       {\cal R}^I_L (y)=(1-{\rm e}^{-\gamma r_{Nd}})^{L+1}G_L(y)
\label{eq:reg}
\end{equation}
where $r_{Nd}=(\sqrt{3}/2)\;y$ is the nucleon-deuteron separation and the
parameter $\gamma$ is fixed requiring that ${\cal R}^I_L (y)\equiv G_L(y)$
asymptotically. Moreover, $G_L(y)$ is
the irregular Bessel function or the irregular Coulomb function in the case of
$n-d$ or $p-d$ scattering, respectively. The description for $A=4$ can be found
in Ref.~\cite{fisher06}

A general
three- or four-nucleon scattering wave function for an incident
state with relative orbital angular momentum $L$, spin $S$, total angular momentum
$J$ and energy below the three-particle breakup threshold is
\begin{equation}
 |\Psi^+_{LSJ}>=\sum_{\alpha,[K],m}A^{LSJ}_{\alpha,[K],m}|\alpha,[K],m>+|\Omega^+_{LSJ}>\ ,
\end{equation}
and its complex conjugate is $\Psi^-_{LSJ}$. A variational estimate of the
trial parameters in the wave function $\Psi^+_{LSJ}$ can
be obtained by requiring, in accordance with
the generalized KVP, that the functional
\begin{equation}
[{}^J{\cal L}^{SS'}_{LL'}]= {}^J{\cal L}^{SS'}_{LL'}-{2\over {\rm det}(u)}
\langle\Psi^-_{LSJ}|H-E|\Psi^+_{L'S'J}\rangle \ ,
\label{eq:kohn}
\end{equation}
be stationary.  Applications of the complex KVP for $N-d$ scattering 
can be found for example in Refs.~\cite{kiev97,kiev01a,marcucci09}.
In the case in which the variational principle is formulated 
in terms of the ${\cal K}$-matrix, we get:
\begin{equation}
[{}^J{\cal K}^{SS'}_{LL'}]= {}^J{\cal K}^{SS'}_{LL'}-
\langle\Psi^-_{LSJ}|H-E|\Psi^+_{L'S'J}\rangle \ .
\label{eq:ckohn}
\end{equation}
Calling the set of indeces $\mu\{\equiv\alpha,[K],m\}$ and $i=\{L,S,J\}$,
the variation of the functional $[\;{}^J{\cal K}^{SS}_{LL}]\equiv[\;{\cal K}_{ii}]$ 
with respect
to the linear parameters $A^i_\mu$ leads to the following two sets of linear equations
\bea
   \label{eq:set1}
  \sum_{\mu'}<\mu|H-E|\mu'>A^{0,i}_{\mu'}=-<\mu|H-E|\Omega^0_i>  \\
  \sum_{\mu'}<\mu|H-E|\mu'>A^{1,i}_{\mu'}=-<\mu|H-E|\Omega^1_i> \ ,
\eea
in accordance of the two possible asymptotic scattering states
$\Omega^0_i$ and $\Omega^1_i$. From the above equations the two sets
of coefficients $A^{0,i}_\mu,A^{1,i}_\mu$ can be obtained. Furthermore, multiplying
the sets by these coefficients and summing on $\mu$, it is possible to reconstruct
the scattering state and the above equations can be formally cast as
\be
   \label{eq:set2}
  <\Psi_C|H-E|\Psi^+_i>=0 \, .
\ee

The variation of the functional with respect to the linear parameters ${\cal K}_{ij}$
results
\be
   \label{eq:var1}
  \delta_{ij}- <\Omega^1_j|H-E|\Psi^+_i> - <\Psi^-_i|H-E|\Omega^1_j>=0 \, .
\ee
Using the normalization condition
\be
   \label{eq:norm1}
  <\Omega^0_i|H-E|\Omega^1_j> - <\Omega^1_j|H-E|\Omega^0_i> = \delta_{ij} \; ,
\ee
the scattering wave function verifies
\bea
   \label{eq:norm2}
  <\Psi^-_i|H-E|\Omega^1_j> - <\Omega^1_j|H-E|\Psi^+_i> = \delta_{ij}  \\
  <\Omega^0_i|H-E|\Psi^+_j> - <\Psi^-_j|H-E|\Omega^0_i> = {\cal K}_{ij} \; ,
   \label{eq:norm3}
\eea
allowing to reduce Eq.(\ref{eq:var1}) to
\be
   \label{eq:var2}
  <\Omega^1_j|H-E|\Psi^+_i>=0 \, .
\ee

The second order estimates of the ${\cal K}$-matrix elements
$[\;^J{\cal K}^{SS'}_{LL'}]\equiv[\;{\cal K}_{ii'}]$ are obtained replacing in the
functional of Eq.(\ref{eq:ckohn}), the first order solutions given by
Eqs.(\ref{eq:set2}) and (\ref{eq:var2}). It results
\begin{equation}
\label{eq:ckohn1}
[\;{\cal K}_{ii'}] = \;{\cal K}_{ii'} - <\Omega^0_i|H-E|\Psi^+_{i'}>
\end{equation}
that can be further reduced using Eq.(\ref{eq:norm3}) to
\begin{equation}
\label{eq:ckohn2}
[\;{\cal K}_{ii'}] = -<\Psi^-_{i}|H-E|\Omega^0_{i'}> \; .
\end{equation}
This final form of the KVP is a direct consequence of the particular
form selected for the asymptotic scattering state given in Eq.(\ref{eq:scata})
in which the flux of the regular wave $\Omega^0_i$ has been set to one.
As we will see in the following, it is useful to define an asymptotic 
scattering state with general
coefficients in both the regular and irregular waves. Accordingly,
the asymptotic scattering state  now reads
\begin{equation}
\Omega^+_{LSJ} =  \sum_{i=1,3}\left[ 
 \sum_{L'S'}{}^J\!\!{\cal A}^{SS'}_{LL'}\,\Omega^0_{L'S'J}(\x_i,\y_i)+
 \sum_{L'S'}{}^J{\cal B}^{SS'}_{LL'}\,\Omega^1_{L'S'J}(\x_i,\y_i) \right] \ .
\label{eq:scatb}
\end{equation}
The coefficients ${}^J\!\!{\cal A}^{SS'}_{LL'}$ and ${}^J{\cal B}^{SS'}_{LL'}$
form the matrices ${\cal A}$ and ${\cal B}$ respectively and the scattering
matrix results ${\cal K}={\cal A}^{-1}{\cal B}$. Starting with a scattering state
that has this asymptotic behavior, the relations of Eqs.(\ref{eq:norm2}) and
(\ref{eq:norm3}) result
\bea
   \label{eq:norm4}
  <\Psi^-_i|H-E|\Omega^1_j> - <\Omega^1_j|H-E|\Psi^+_i> = {\cal A}_{ij}  \cr
  <\Omega^0_i|H-E|\Psi^+_j> - <\Psi^-_j|H-E|\Omega^0_i> = {\cal B}_{ij} \; ,
   \label{eq:norm5}
\eea
and, using Eqs.(\ref{eq:var2})-(\ref{eq:ckohn2}), the KVP takes the particular form
\bea
{\cal A}_{ij}&=&<\Psi^-_i|H-E|\Omega^1_j>   \cr
[{\cal B}_{ij}]&=&-<\Psi^-_i|H-E|\Omega^0_j> \cr
[{\cal K}]&=&{\cal A}^{-1}[{\cal B}] 
   \label{eq:ir1}
\eea
where $[{\cal B}]$ and $[{\cal K}]$ are second order estimates. 

Eqs.(\ref{eq:ir1}) formulate the KVP in terms of integral relations depending
on the internal structure of the scattering wave function $\Psi^-_i$. In fact
$(H-E)\Omega^0_j$ and $(H-E)\Omega^1_j$ go to zero as each of the three
Jacobi coordinates $y_k$ goes to $\infty$ $(k=1,2,3)$, since
$\Omega^{0,1}_j$ are the solutions of $(H-E)$ in that limit. Therefore,
in Eqs.(\ref{eq:ir1}), it would be possible to use trial wave functions $\Psi^-_i$
that are solutions of $(H-E)$ in the interaction region but do not have the physical
asymptotic behavior indicated in Eq.(\ref{eq:scatb}). In particular, it would be
possible to use the bound-state wave functions $\Psi_n$ described in the previous 
section to calculate the scattering matrix corresponding to a center-of-mass
energy $E^0_n$. This is discussed in the next section.

\section{Scattering matrix from bound-state-like wave functions}

\begin{figure}[h]
\includegraphics[width=16cm]{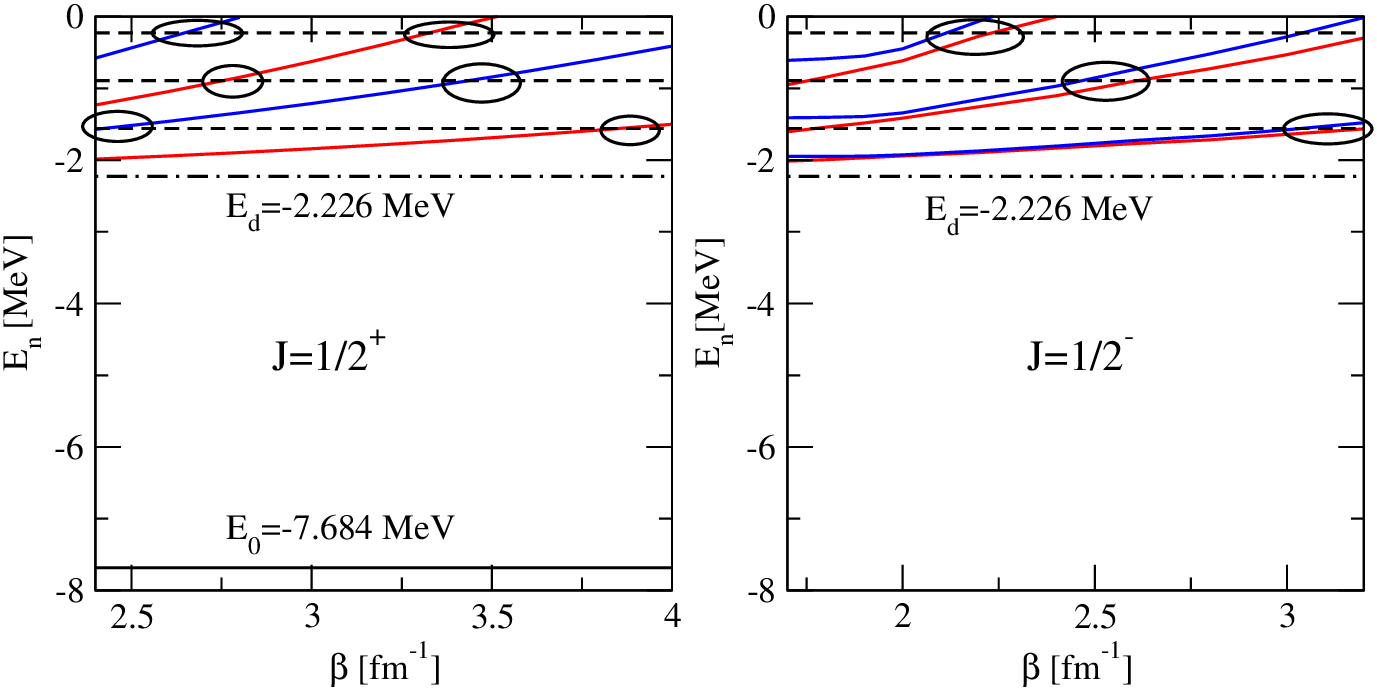}
\caption{(Color online)
The same as Fig.\protect\ref{fig:fig1} in which two sequential solutions 
having the same eigenvalue are selected (indicated by the circles) in the three
cases corresponding to incident energies $E_{lab}=1,2,3$ MeV respectively.} 
\label{fig:fig2}
\end{figure}

In section II the construction of $A=3,4$ bound states having general
quantum numbers $J^\pi$ corresponding to different levels with negative
eigenvalues $E_n$ has been discussed. In the case of the $A=3$ state $J=1/2^+$ 
the $E_0$ level and the
corresponding wave function $\Psi_0$ describe the energy and structure of the 
triton or $^3$He for the two possible values of $T_z=1/2$ or $-1/2$, respectively. Using
the nonlinear parameter $\beta$ as a control parameter it was possible to
construct states with eigenvalues $E_n$ in the region $E_d<E_n<0$.
In a similar way, it is possible to contruct these kind of states for
arbiratry values of $J^\pi$. As an example, in section II, the case $J=1/2^-$
was explicitly discussed. Furthermore, it was shown that these states organize
sequentially having occupation probabilities that can be connected with the
different components of a scattering state, corresponding to the different
values of the quantum numbers $L,S,J$. The number of these components fixes
the dimension of the scattering matrix and, correspondingly, the degeneration of
the state. Therefore, in order to construct a scattering state using 
bound-state-like functions, those components can be taken into account considering
sequential solutions having the same energy. To this end the control
parameter $\beta$ can be used to select sequential solutions at the same
eigenvalue $E_n$. This is shown in Fig.~\ref{fig:fig2} for three different cases.
The three dashed lines in both panels of the figure indicate the energies
corresponding to incident energies in the lab system $E_{lab}=1,2,3$ MeV. 
As explained in section II, the red and blue lines show the variation of
sequential eigenvalues as a function of $\beta$ having the different
structures given in Table~\ref{tb:tab1}. The circles in Fig.~\ref{fig:fig2}
indicate the points in wich the eigenvalues cross the dashed lines and, accordingly,
at those specific values of $\beta$ two different solutions, $\Psi^1_n$ and 
$\Psi^2_n$, can be constructed having the same energy $E_n$ and presenting
a very different internal structure. These two states are solutions of
$(H-E_n)\Psi^{1,2}_n=0$ in the internal region and, since they are square
integrable states, they go to zero asymptotically. However the integral relations
of Eq.(\ref{eq:ir1}) depend on the internal part of the wave function and
therefore it would be possible to use $\Psi^1_n$ and $\Psi^2_n$ as
trial wave functions. In this case the second order estimate of the scattering
matrix is
\bea
{\cal A}_{ij}&=&<\Psi^i_n|H-E_n|\Omega^1_j>   \cr
[{\cal B}_{ij}]&=&-<\Psi^i_n|H-E_n|\Omega^0_j> \cr
[{\cal K}]&=&{\cal A}^{-1}[{\cal B}] 
   \label{eq:ir2}
\eea
where $i,j$ indicate either the two solutions, $\Psi^{1,2}_n$, and the two
possible values of the set of quantum numbers $(L,S,J)$ in $J=1/2^{\pm}$.

\begin{figure}[h]
\includegraphics[width=16cm]{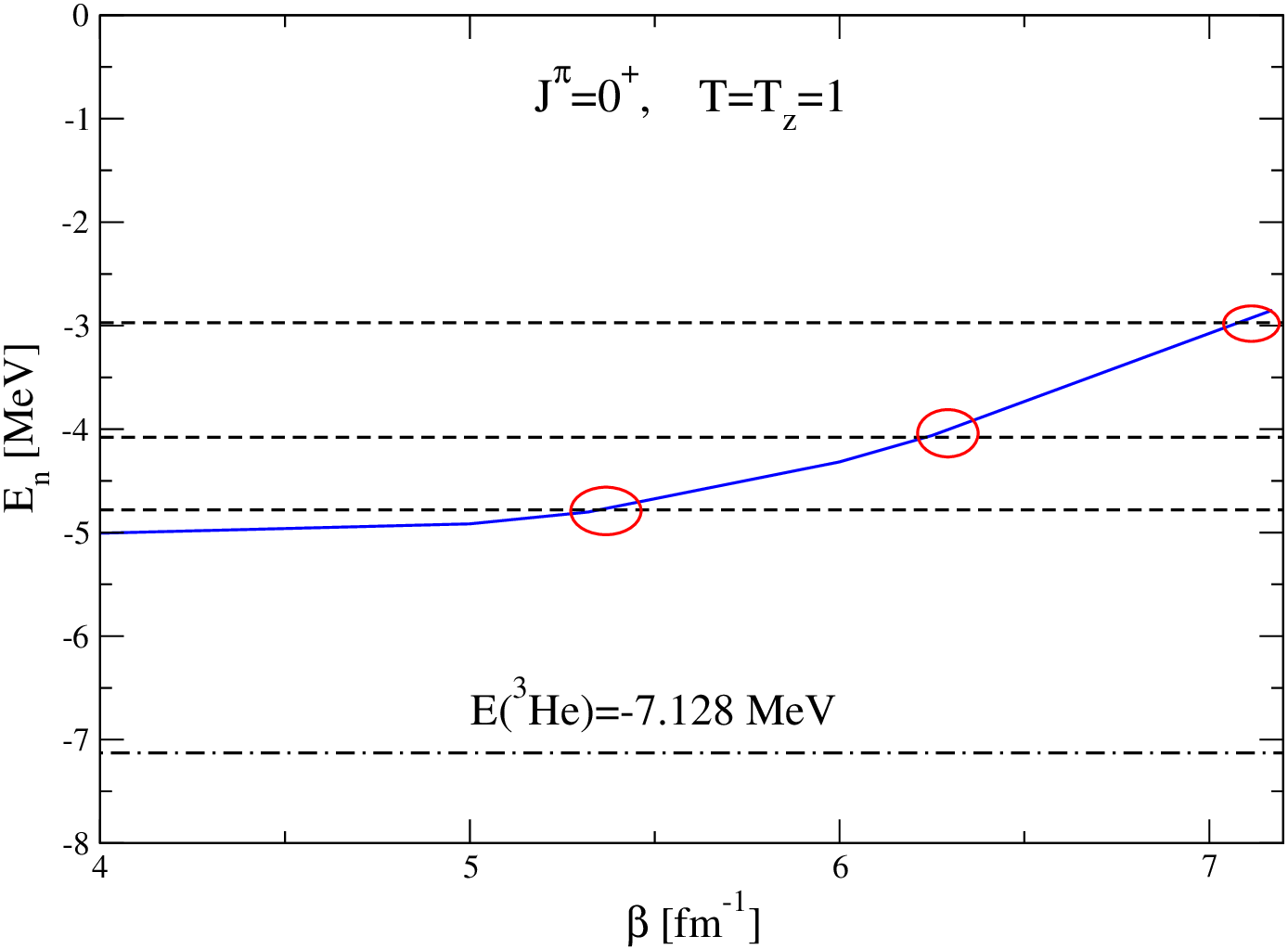}
\caption{(Color online)
The same as Fig.\protect\ref{fig:fig4b1} in which the lowest
eigenvalue is selected (indicated by the circles) in the three
cases corresponding to the lab energies $E_{cm}=3.13,4.05,5.54$ MeV respectively.} 
\label{fig:fig4b2}
\end{figure}

For the $A=4$ case we have analyzed the single channel $J^\pi=0^+$ state with
$T=T_z=1$. In Fig.~\ref{fig:fig4b2} we show the three cases (indicated with circles)
at which, for specific values of the control parameter $\beta$, the eigenvalue
matches the selected energies. Accordingly the second order estimate
of the scattering matrix can be obtained in each case as
\bea
{\cal A}&=&<\Psi_n|H-E_n|\Omega^1>   \cr
[{\cal B}]&=&-<\Psi_n|H-E_n|\Omega^0> \cr
[{\cal K}]&=&{\cal A}^{-1}[{\cal B}]
   \label{eq:ir3}
\eea
In this case we are considering a single channel state and therefore the scattering
matrix results a scalar. 

\begin{figure}[h]
\includegraphics[width=16cm]{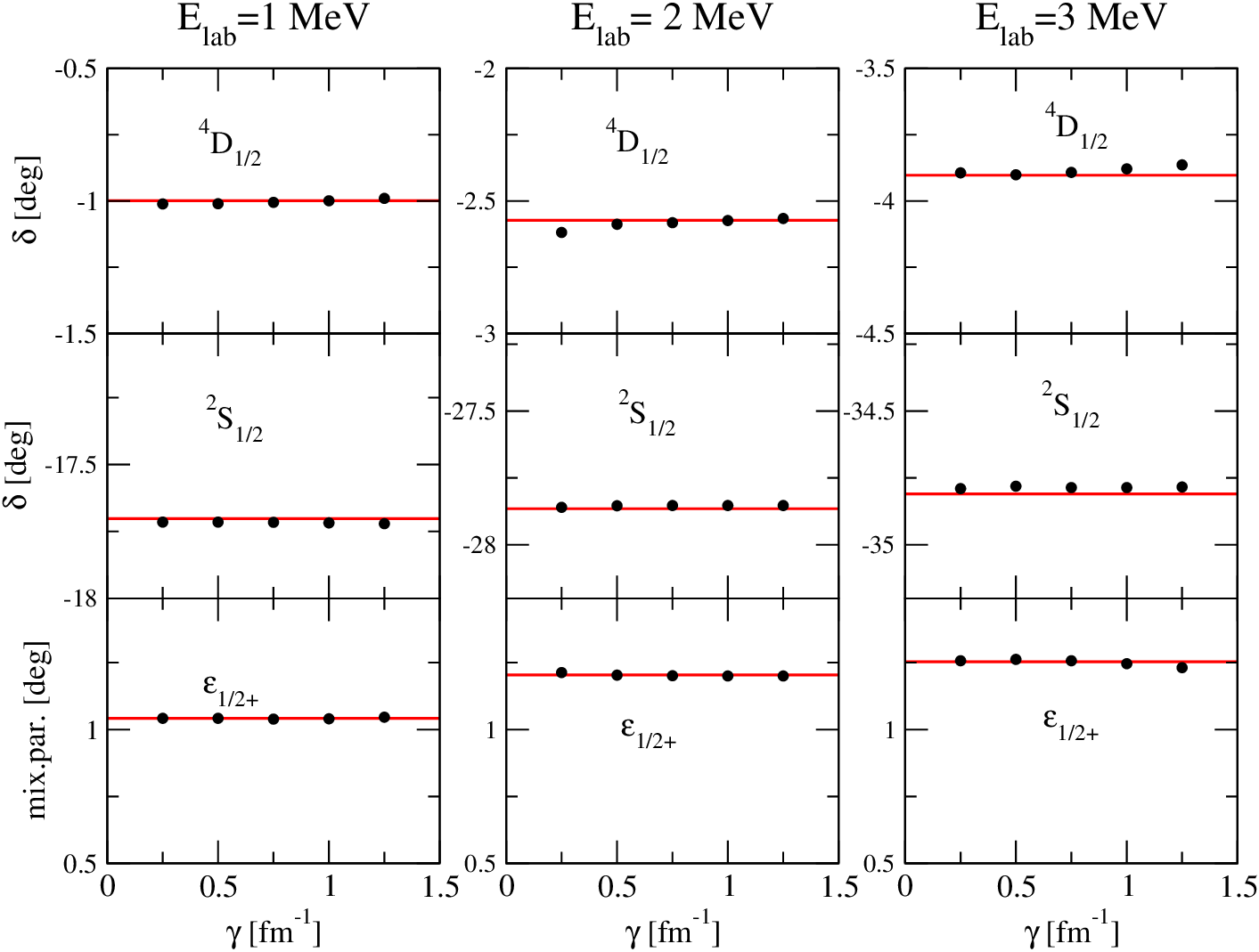}
\caption{(Color online)
The $n-d$ $J^\pi=1/2^+$ phase-shifts and mixing parameters as a function of the
regularization parameter $\gamma$ at the three indicated energies.
The red line corresponds to the results of Ref.~\protect\cite{benchmark1}.}
\label{fig:fig3}
\end{figure}

\begin{figure}[h]
\includegraphics[width=16cm]{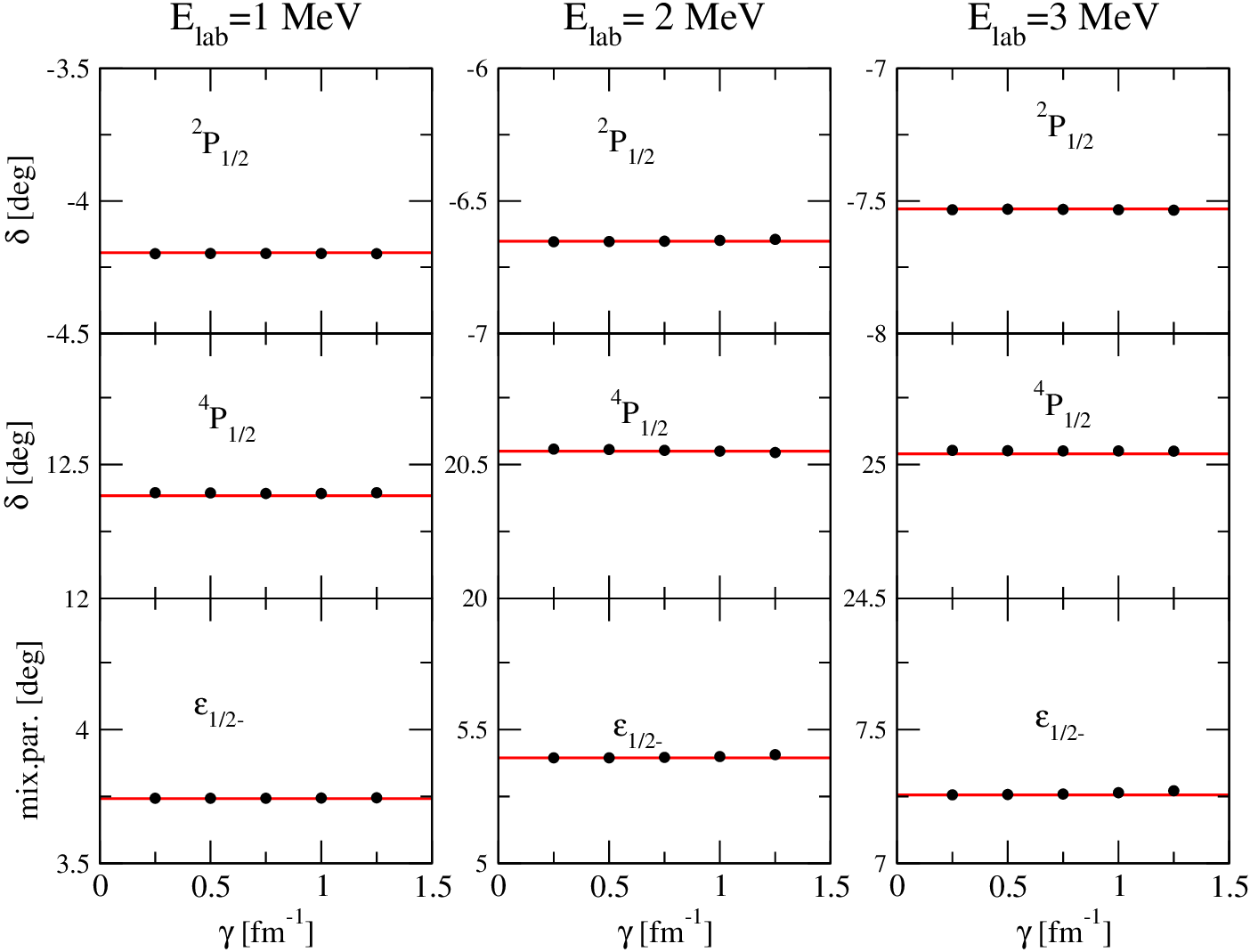}
\caption{(Color online)
The $n-d$ $J=1/2^-$ phase-shifts and mixing parameters, for the AV14 potential, 
as a function of the regularization parameter $\gamma$ at the three indicated 
energies. The red line corresponds to the results of Ref.~\protect\cite{benchmark1}.}
\label{fig:fig4}
\end{figure}

In the following, results of phase-shifts and mixing parameters for the
$n-d$ system, calculated using the AV14 $NN$ potential, 
are presented for the $J^\pi=1/2^+$ state in Fig.~\ref{fig:fig3}, and
for the $J^\pi=1/2^-$ state in Fig.~\ref{fig:fig4}, 
at the three selected energies $E_{lab}=1,2,3$ MeV.
The stability of the results with $\gamma$, the regularization
parameter introduced in Eq.(\ref{eq:reg}), is chosen as a convergence criterion.
This criterion has been discussed in Refs.~\cite{intrel,kiev10a} and essentially
it establishes the quality of $\Psi_n^i$ as solution of $(H-E_n)\Psi_n^i=0$. 
In fact, if $\Psi_n^i$ is a good solution, the integrals of Eq.(\ref{eq:ir2})
are largely independent of $\gamma$.
The results are
compared to the benchmark of Ref.~\cite{benchmark1} given in the figures as a red
line. The results of the application of Eq.(\ref{eq:ir2}) are shown as filled
circles corresponding to values of $\gamma$ varying from $0.25$ fm$^{-1}$ to
$1.25$ fm$^{-1}$. We can observed a good stability on this interval and, furthermore,
the results are in very good agreement with those of Ref~\cite{benchmark1}.

\begin{figure}[h]
\includegraphics[width=16cm]{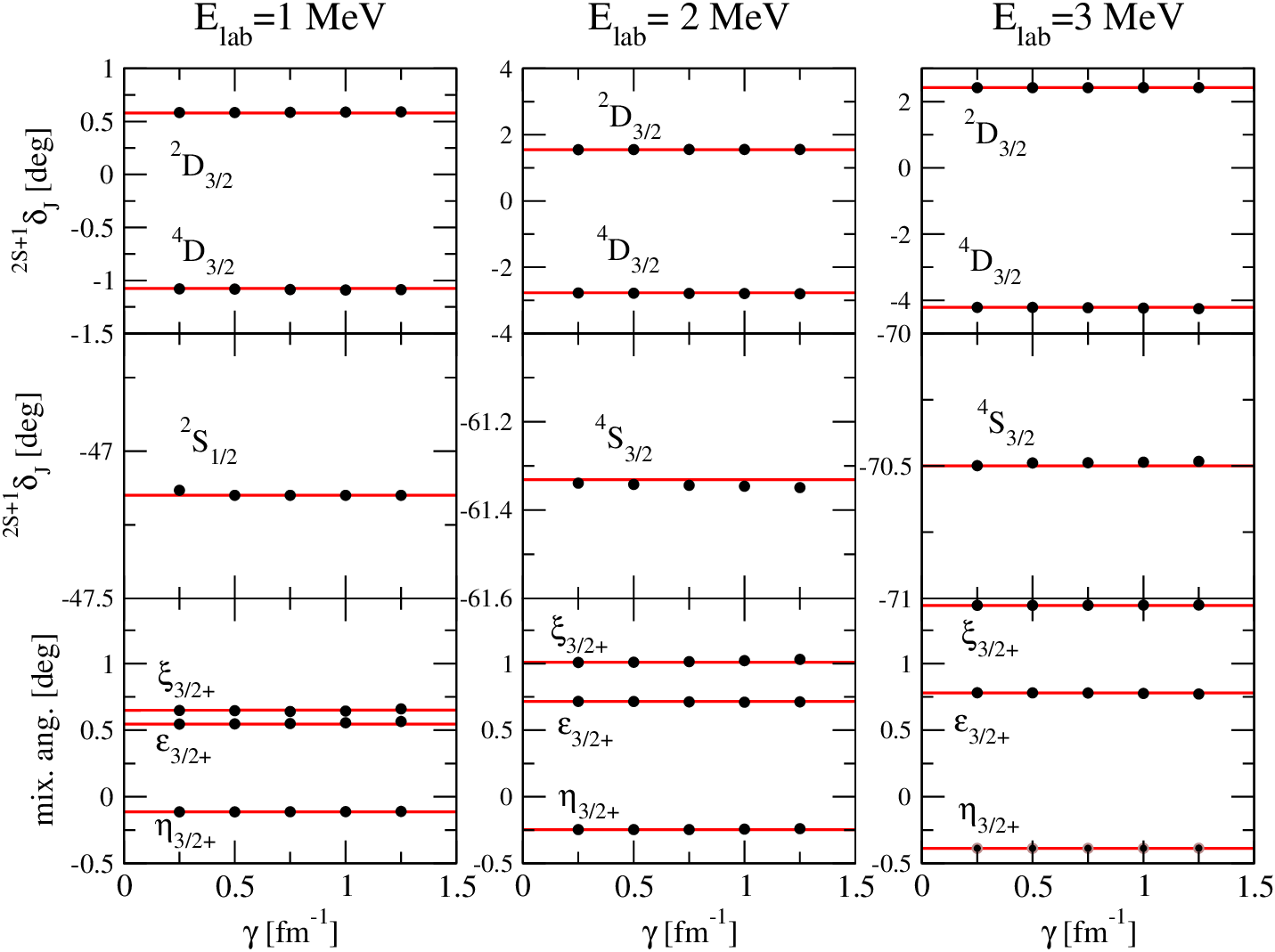}
\caption{(Color online)
The $n-d$ $J=3/2^+$ phase-shifts and mixing parameters, for the AV14 potential, 
as a function of the regularization parameter $\gamma$ at the three indicated energies.
The red line corresponds to the results of Ref.~\protect\cite{benchmark1}.}
\label{fig:fig5}
\end{figure}

\begin{figure}[h]
\includegraphics[width=16cm]{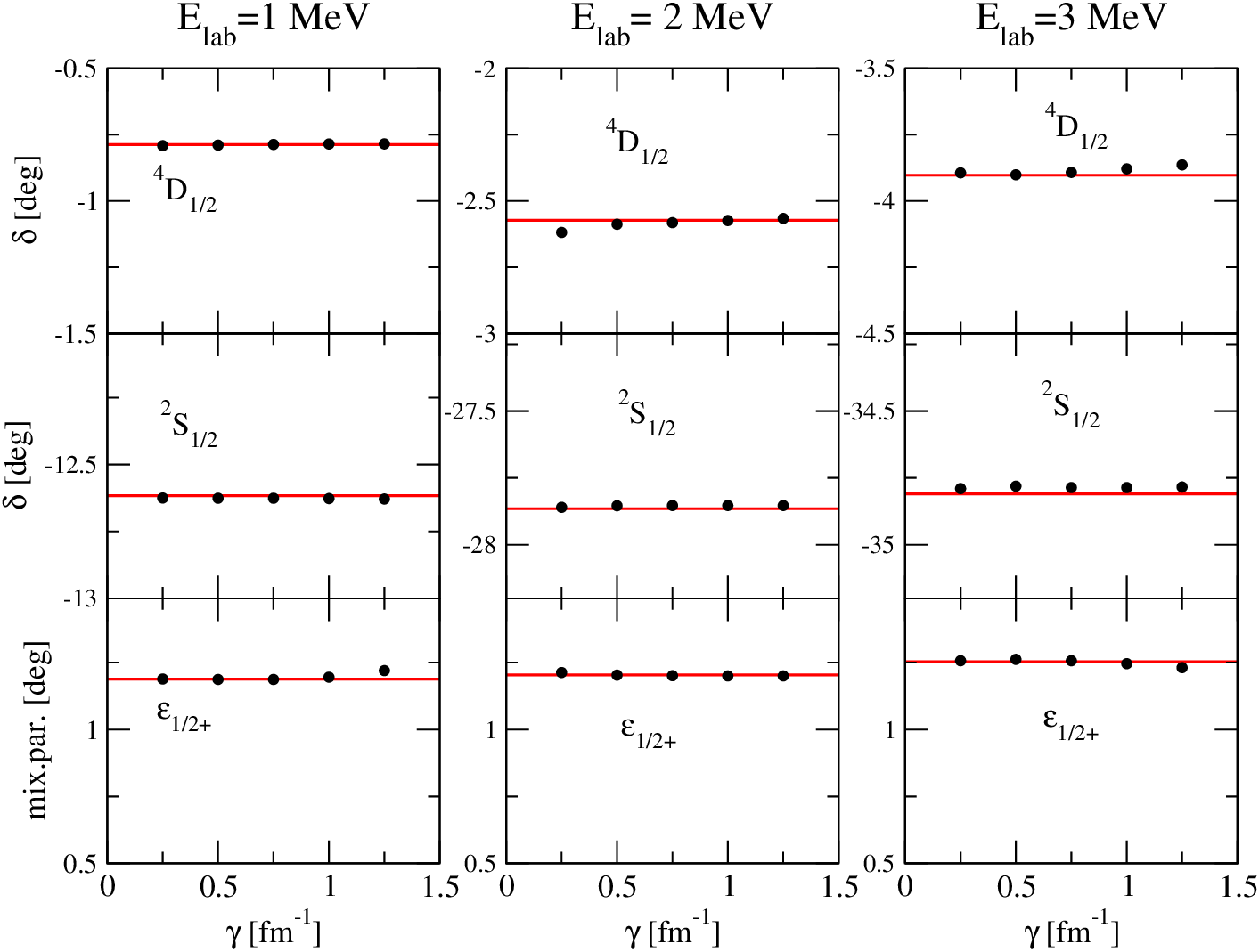}
\caption{(Color online)
The $p-d$ $J=1/2^+$ phase-shifts and mixing parameters, for the AV14 potential, 
as a function of the regularization parameter $\gamma$ at the three indicated energies.
The red line corresponds to the results of Ref.~\protect\cite{kiev01b}.}
\label{fig:fig6}
\end{figure}

In Fig.~\ref{fig:fig5} results are given for the $n-d$ $J^\pi=3/2^+$ state. In this case
the ${\cal K}$-matrix is a $3\times 3$ matrix, corresponding to asymptotic
configurations having $L=0,S=3/2$, $L=2,S=1/2$ and $L=2,S=3/2$. The diagonalization
of the Hamiltonian matrix in the $J^\pi=3/2^+$ case
produces sequential eigenvalues with occupation probabilities
in accordance with these three configuration. Using the control parameter
$\beta$, three sequential eigenvalues can be chosen to have a particular $E_n$
value as has been done before for the $J^\pi=1/2^{\pm}$ states. In the figure we observe
a good stability with the regularization parameter $\gamma$ and a close
agreement with the results of the Ref.~\cite{benchmark1}. 
In Fig.~\ref{fig:fig6} results for the
$p-d$ $J=1/2^+$ state are given. The description of the $p-d$ process at low
energies presents some problems using the Faddeev equations. In Ref.~\cite{kiev01b}
a benchmark for $p-d$ scattering has been produced using the HH method and
the Faddeev method in configuration space. The results of the benchmark
are shown as a red line in Fig.~\ref{fig:fig6}. From the figure we can observe that
the results using the integral relations reproduce extremely well the benchmark
results. This is an important point since in bound-state-type calculations 
the treatment of the Coulomb potential does not present any troubles.

\begin{figure}[h]
\includegraphics[width=16cm]{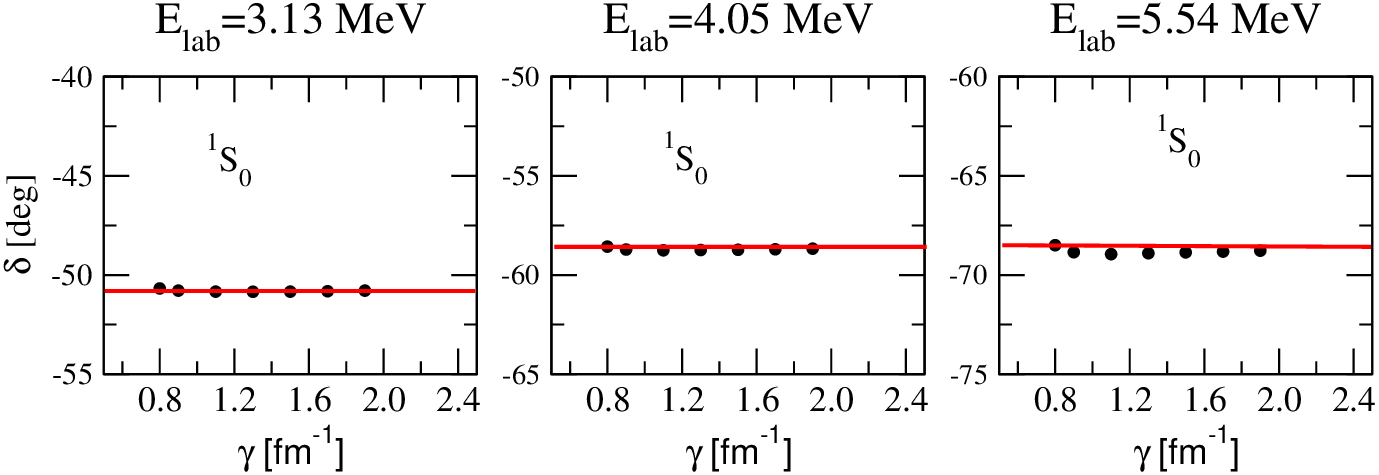}
\caption{(Color online)
The $p-^3\!{\rm He}$ $J=0^+$ phase-shift, for the N3LO potential, as a function of the
regularization parameter $\gamma$ at the three indicated energies.
The red line corresponds to the results of Ref.~\protect\cite{benchmark4b}.}
\label{fig:fig4b3}
\end{figure}

The $p-^3{\rm He}$ results are given in Fig.~\ref{fig:fig4b3} for the three selected
energies. The phase-shift for the $0^+$ state is shown as a function of the
regularization parameter $\gamma$ (filled circles). As a comparison, the results of the 
recent benchmark of Ref.~\cite{benchmark4b} are shown as a red line. We can observe a good
stability with $\gamma$ indicating that the four-nucleon bound-state eigenfunction 
$\Psi$ is a good solution of $(H-E)\Psi=0$ at the specified energies. Moreover
the results obtained using the integral relations are in close agreement with those of 
the benchmark.

\section{Conclusions}

In this work the elastic scattering matrix has been determined using 
bound-state-like wave functions. To this end two integral relations derived from the
KVP have been used. Initially, these integral relations were derived 
in Ref.~\cite{intrel} in order to extract phase shifts from the
solutions calculated using the hyperspherical adiabatic expansion in the three-nucleon
system. In this method the boundary conditions at large distances are imposed
in terms of the hyperradius $\rho$. However, as explained in Ref.~\cite{intrel},
the boundary conditions depend explicitly on the Jacobi 
coordinates $\x_i,\y_i$ describing the asymptotic configuration of a
deuteron formed by particles $(j,k)$ and an incoming nucleon (particle $i$).
The equivalence between imposing the boundary conditions in $\rho$ or in the Jacobi coordinates
directly, results at very large values of the hyperradius where the relation 
$\rho\approx y_i$ is verified. As a consequence, the phase shifts obtained from the adiabatic 
expansion require a large number of terms to converge. On the other hand, 
the phase shifts obtained as a quotient of the two integral relations
converge much faster and, in fact, the rate of convergence is
similar to that one obtained in the case of bound state solutions.
The reason behind this fact is that the integral relations depend only on the
internal part of the wave function. Therefore, it is enough that the wave
function verifies $(H-E)\Psi=0$ in the internal region to obtain almost exact
results for the scattering matrix at the center-of-mass energy $E$. 
This characteristic allows to apply the integral relations using
bound-state-like wave functions obtained from a direct diagonalization of the
Hamiltonian $H$.
Eigenvectors corresponding to eigenvalues belonging to the continuum spectrum of
$H$ can be used as inputs to determine the scattering matrix at fixed values of
$E$. Applications for single-channel solutions using semi-realistic
$NN$ potentials are given in Ref.~\cite{kiev10a}. The coupled-channel case of an atom
colliding a dimer formed by other two atoms is given in Ref.~\cite{romero10}.

In the present work, applications to elastic scattering of a nucleon on a deuteron ($A=3$)
or on $^3$He ($A=4$) below the breakup threshold, using realistic nucleon-nucleon potentials 
has been discussed. In particular, for $A=3$, two or three solutions at the same
energy have to be determined corresponding to the different possible asymptotic
configurations of the system. A detailed construction of such solutions, using the 
nonlinear parameter $\beta$ as a control parameter, has been analyzed. 
Moreover it has been shown that the eigenvectors of successive
eigenvalues organize in pairs (for $J=1/2^{\pm}$) or in triplets (for $J>1/2$),
corresponding to the different asymptotic structures. The control parameter
$\beta$ has been tuned to find solutions having the same eigenvalue that have been
usend to calculate the scattering matrix at the selected energy. The obtained results are 
in close agreement with those presented in the $A=3$ benchmarks of 
Refs.~\cite{benchmark1,kiev01b} and in the $A=4$ benchmark of Ref.\cite{benchmark4b}. 
In particular, the results
for $p-d$ and $p-^3{\rm He}$ scattering demonstrate that the scattering matrix
can be calculated using bound-state-like wave functions also in scattering of charged particles. 

Well established bound-state methods to diagonalize the nuclear Hamiltonian 
in systems with $A>4$ already exist. The formulation of the scattering
matrix presented in this work will allow to extent these studies to the low-energy
continuum spectrum. It will be then possible to compare 
theoretical predictions for scattering observables to the experimental data,
in order to evaluate the capability of the present models for the 
interaction to describe the nuclear structure. 
Studies along this line are at present intensively pursued.

\newpage

\begin{table}[h]
\begin{tabular}{cccc}
\hline
 $J=1/2^+$ & $P_S(\%)$ & $P_P(\%)$ & $P_D(\%)$ \cr
$E_0$      & 90.96     & 0.08      &  8.97     \cr
\hline
$E_1$      & 90.04     & 0.00      &  5.96      \cr
$E_2$      &  1.22     & 2.72      &  96.06     \cr
\hline
$E_3$      & 94.20     & 0.00      &  5.80      \cr
$E_4$      &  1.21     & 2.70      &  96.09     \cr
\hline
\hline
 $J=1/2^-$ & $P^{1/2}_P(\%)$ & $P^{3/2}_P(\%)$ & $P_D(\%)$ \cr
$E_1$      &  3.36     & 94.87      &  1.77     \cr
$E_2$      & 93.80     & 3.42      &  2.78      \cr
\hline
$E_3$      &  3.47     & 94.71      &  1.82     \cr
$E_4$      & 93.42     & 3.86      &  2.72      \cr
\hline
$E_5$      &  4.32     & 93.87      &  2.01     \cr
$E_6$      & 92.72     & 4.77      &  2.51      \cr
\hline
\end{tabular}
\caption{Occupation probabilities of the different states shown in
Fig.~\protect\ref{fig:fig1} at $\beta=2.5$ fm$^{-1}$
($J^\pi=1/2^+$) and $\beta=2.0$ fm$^{-1}$ ($J^\pi=1/2^-$).}
\label{tb:tab1}
\end{table}

\newpage 

\end{document}